\documentclass[10pt,twocolumn,letterpaper]{article}

\usepackage[pagenumbers]{cvpr} 

\usepackage{algorithmic}
\usepackage{graphicx}
\usepackage{textcomp}
\usepackage{xcolor}
\usepackage{multirow}

\usepackage{footnote} 
\usepackage[]{siunitx}
\DeclareSIUnit{\degree}{°}
\DeclareSIUnit{\deg}{deg}
\DeclareSIUnit{\nothing}{\relax}
\DeclareSIUnit\pixel{px}
\DeclareSIUnit{\op}{Op}
\DeclareSIUnit{\fps}{frame/s}

\usepackage{color}
\usepackage{tabularray}
\usepackage{array}
\usepackage{booktabs}
\usepackage{makecell}
\usepackage{threeparttable}

\usepackage{pifont} 
\newcommand{\cmark}{\ding{51}} 

\definecolor{cvprblue}{rgb}{0.21,0.49,0.74}
\usepackage[pagebackref,breaklinks,colorlinks,allcolors=cvprblue]{hyperref}

\usepackage[accsupp]{axessibility}  

\def\paperID{7} 
\def\confName{CVPR}
\def\confYear{2026}

\titleheader{\darkgrayed{This paper has been accepted for publication at the IEEE Conference on Computer Vision and Pattern Recognition (CVPR) Workshops, 2026
\copyright IEEE}}

\title{TinyDEVO: Deep Event-based Visual Odometry on Ultra-low-power \\ Multi-core Microcontrollers}

\author{
    \thanks{Both authors contributed equally.} Alessandro Marchei$^1$
    \quad
    $^{*}$Lorenzo Lamberti$^{1,2}$
    \quad
    Daniele Palossi$^{1,2}$ 
    \quad
    Luca Benini$^{1,3}$\\
    $^1$IIS, ETH Z\"urich 
    \quad
    $^2$IDSIA, USI-SUPSI 
    \quad
    $^3$DEI, University of Bologna\\
}

\begin{document}
\maketitle

\begin{abstract}
A key task in embedded vision is visual odometry (VO), which estimates camera motion from visual sensors, and it is a core component in many embedded power-constrained systems, from autonomous robots to augmented/virtual reality wearable devices.
The newest class of VO systems combines deep learning models with bio-inspired event-based cameras, which are robust to motion blur and lighting conditions.
However, State-of-the-Art (SoA) event-based VO algorithms require significant memory and computation, e.g., the SoA-leading DEVO requires \SI{733}{\mega\byte} and \SI{155}{\giga\nothing } multiply-accumulate (MAC) operations per frame. 
We present TinyDEVO, an event-based VO deep learning model designed for resource-constrained microcontroller units (MCUs).
We deploy TinyDEVO on an ultra-low-power (ULP) 9-cores RISC-V-based MCU, achieving a throughput of $\sim$\SI{1.2}{frame/\second} with an average power consumption of only \SI{86}{\milli\watt}.
Thanks to our neural network architectural optimizations and hyperparameter tuning, TinyDEVO reduces the memory footprint by 11.5$\times$ (to \SI{63.8}{\mega\byte}) and the number of operations per frame by 29.7$\times$ (to \SI{5.2}{\giga MAC/frame}) w.r.t. DEVO, while maintaining an average trajectory error of \SI{27}{\centi\meter}, i.e., only \SI{19}{\centi\meter} higher than DEVO, on three SoA datasets.
Our work demonstrates, for the first time, the feasibility of an event-based VO pipeline on ULP devices.
\end{abstract}

\section*{Supplementary Video}
\url{https://youtu.be/wUx0V9psvUk}

\section{Introduction} \label{sec:introduction}


\begin{table*}[t]
\centering
\small
\caption{
Frame/Event monocular VO pipelines overview, either based on a geometric (GEO) algorithm or a deep learning-based (DL) one. Memory is reported as peak memory allocation.
N.D. means not declared by the authors. 
}
\label{tab:related_works}
\begin{threeparttable}
\begin{tabular}{lccccccc}
\toprule
\textbf{Work} & \textbf{Type} & \textbf{Input} & \textbf{Resolution} & \textbf{Memory [MB]} & \textbf{Device} & \textbf{FPS / Event-rate} & \textbf{Power [W]} \\ 
\midrule
ORB-SLAM3~\cite{orb_slam3,legittimoBenchmarkAnalysisDatadriven2023} & GEO & frame & 752×480 & 900 & Jetson Xavier AGX & 23.9~FPS & 30 \\
PackNet~\cite{legittimoBenchmarkAnalysisDatadriven2023} & DL & frame & 752×480 & 3000 & Jetson Xavier AGX & 80~FPS & 30 \\
EVO~\cite{rebecq_evo_2017} & GEO & event & 240×180 & 535* & Intel~i7-4810 & $\sim$1.5~Mevents/s & 47 \\
Ye~\textit{et al.}~\cite{ye2019unsupervisedlearningdenseoptical} & DL & event & 346×260 & N.D. & GTX~1080Ti & 250~FPS & 250 \\
DEVO~\cite{klenk_deep_2023} & DL & event & 240×180 & 733* & RTX~4070 & 27.5~FPS* & 250 \\
\midrule
\multirow{2}{*}{\textbf{TinyDEVO (Ours)}} & \multirow{2}{*}{\textbf{DL}} & \multirow{2}{*}{\textbf{event}} & \multirow{2}{*}{\textbf{240×180}} & \multirow{2}{*}{\textbf{64}} & \textbf{RTX~4070} & \textbf{108 FPS} & \textbf{250} \\
 &  &  &  &  & \textbf{GAP9 SoC} & \textbf{1.2~FPS} & \textbf{0.09} \\
\bottomrule
\end{tabular}
\begin{tablenotes}
  \small
  \item *From our measurements (not provided in the original work).
\end{tablenotes}
\end{threeparttable}
\vspace{-1mm}
\end{table*}

\begin{figure}
    \centering
    \includegraphics[width=1\linewidth]{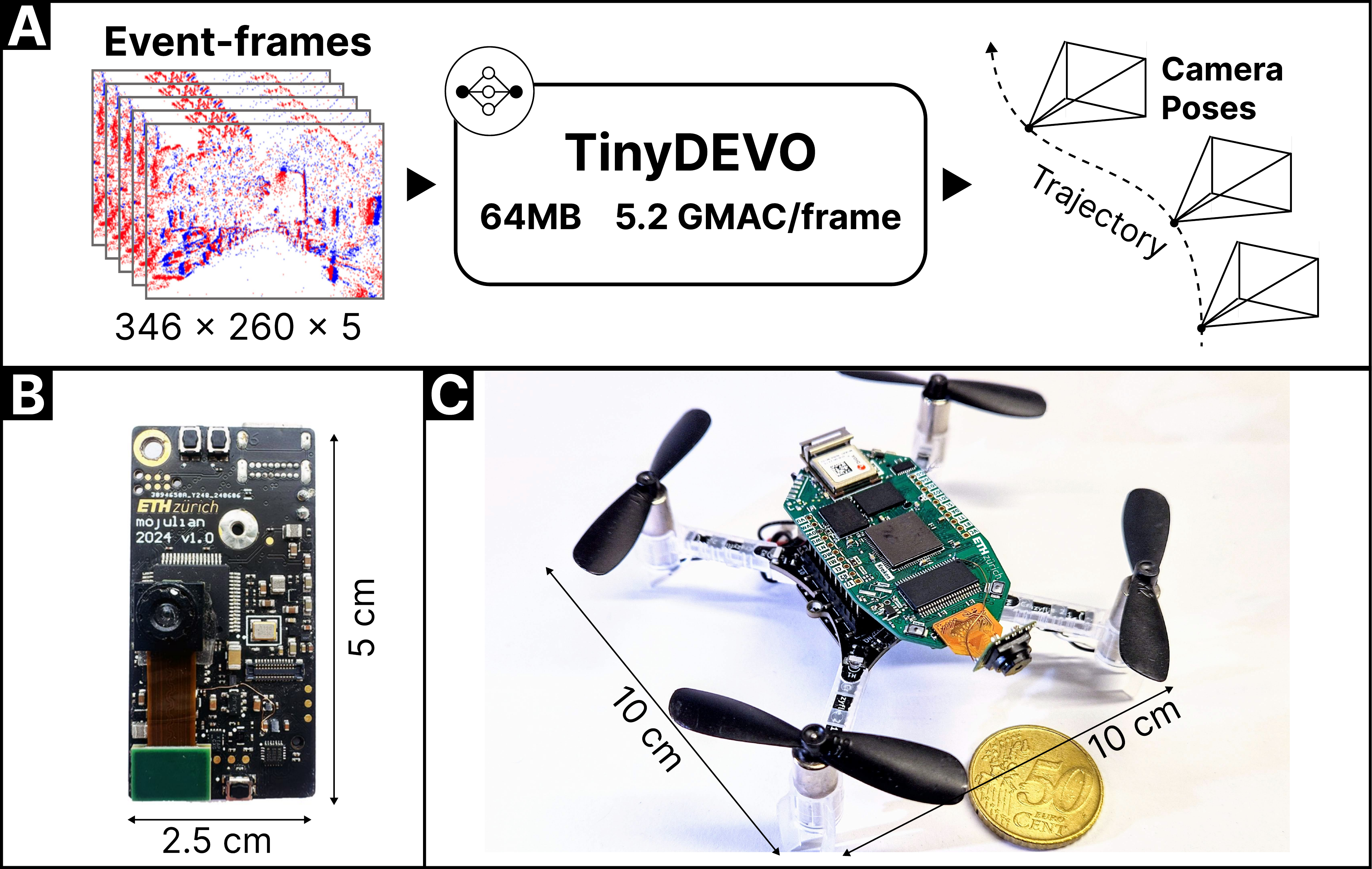}
    \caption{A) TinyDEVO: our DL-based, event-only VO model tailored to embedded vision systems. Examples of embedded platforms using event-based sensing include: B) an IoT wearable device from~\cite{bartoliLynXEventBasedGesture2025}, and C) a miniaturized robot from~\cite{potocnikCircuitsSystemsEmbodied2024}.} 
    \label{fig:intro}
    \vspace{-2mm}
\end{figure}

Embodied artificial intelligence (AI) and agentic AI rely on key fundamental embedded vision tasks, such as monocular visual odometry (VO)~\cite{legittimoBenchmarkAnalysisDatadriven2023}.
Monocular VO estimates the six degrees of freedom of a camera pose from a single visual input (\Cref{fig:intro}-A). 
Originally developed as a core component for perception tasks in large robotic platforms~\cite{teed_deep_2023,niculescu2022fly}, VO has recently become relevant in the edge computing domain employing sub-\SI{100}{\milli\watt} microcontroller units (MCUs)~\cite{palossi_vo} (\Cref{fig:intro}-B-C). 
For instance, VO is essential in smart glasses for augmented and virtual reality~\cite{engelProjectAriaNew2023, Krishnan_2025_ICCV, xr_barbara_salvo, bartoliLynXEventBasedGesture2025} to track the user’s head motion and to ensure that virtual objects are correctly rendered in the user's field of view~\cite{fanEMHIMultimodalEgocentric2025}. 
In robotics, VO provides ego-motion estimation, which is key for full autonomy in tasks such as planning, localization, and mapping~\cite {legittimoBenchmarkAnalysisDatadriven2023,cadena2016past}.
Vision-based motion estimation enables navigation capabilities across a wide spectrum of robotic platforms, spanning from large terrestrial and aerial robots~\cite{faessler2016_vo_drone}, employing power-hungry embedded computers (i.e., 10s of Watts), to miniaturized nano-drones weighing a few tens of grams~\cite{lamberti_pulpdronetv3, suleiman_navion_2019, bouwmeesterNanoFlowNetRealtimeDense2023, potocnikCircuitsSystemsEmbodied2024}, which can host only ultra-low-power (ULP) MCUs.

Event-based cameras~\cite{event_cameras_survey} have recently emerged as a promising bio-inspired sensing technology for enhancing the robustness and accuracy of embedded vision pipelines, including VO ones. 
Unlike traditional frame-based sensors, they capture asynchronous per-pixel brightness changes with microsecond latency and a high dynamic range ($\sim$\SI{140}{\decibel}).
Thanks to their characteristics, event-based sensors enable robust perception even in challenging light conditions, e.g., extremely dark or bright environments, and are robust to motion blur.
Event-based sensors are also power and energy-efficient, with  reported power consumption as low as \SI{10}{\milli\watt}~\cite{event_cameras_survey}. 

Existing VO algorithms can be categorized into geometric and deep learning-based (DL) methods~\cite{legittimoBenchmarkAnalysisDatadriven2023}. 
Geometric methods rely on explicit feature extraction and 3D geometry~\cite{orb_slam3, rebecq_evo_2017, Kim2016RealTime3R}, whereas DL-based pipelines use data-driven representations that achieve higher accuracy, robustness, and generalization~\cite{teed_deep_2023, klenk_deep_2023,legittimoBenchmarkAnalysisDatadriven2023}.
Consequently, DL-based methods now define the State-of-the-Art (SoA) in both frame-based and event-based VO.  
Among them, Deep Event Visual Odometry (DEVO)~\cite{klenk_deep_2023} is the leading monocular event-based pipeline, outperforming frame-based counterparts~\cite{teed_deep_2023} with an average trajectory error (ATE) of \SI{8}{\centi\meter} on 10-\SI{50}{\meter} long trajectories from SoA datasets.
To achieve this result, DEVO requires at least \SI{733}{\mega\byte} of memory and \SI{155}{\giga \nothing} multiply-accumulate (MAC) operations per frame, relying on high-end GPUs such as the Nvidia A40 consuming \SI{250}{\watt}.

In contrast, most of consumer electronic\cite{kuhne_low_2025}, wearable devices~\cite{freyGAPsesVersatileSmart2025, bartoliLynXEventBasedGesture2025}, and miniaturized robots~\cite{lamberti_pulpdronetv3, potocnikCircuitsSystemsEmbodied2024} feature ULP MCUs which provide only a few \SI{}{\mega\byte} of memory and peaks at \SI{150}{\giga MAC/\second} on fixed-precision data workloads~\cite{ceredaTrainingFlyOnDevice2024}.
Our work addresses the challenging scenario of enabling, for the first time, the DEVO full-fledged VO pipeline on an ULP MCU, by presenting our novel DL-based, event-only tiny model for VO.
Our main contributions are:
\begin{enumerate}
    \item leveraging our model size and complexity reduction methodology, we present TinyDEVO, a lightweight event-based VO algorithm tailored to ULP MCUs;
    \item we provide an energy-efficient implementation of TinyDEVO on an ULP multi-core RISC-V MCU, and we profile its end-to-end execution in terms of latency and power consumption;
    \item we present a thorough experimental analysis on the trade-offs between execution performance and VO's accuracy.
\end{enumerate}

As DEVO combines a DL-based feature extractor with a recurrent module to iteratively process features, our workload reduction methodology consists of \textit{i}) \textit{model reduction}, i.e., achieved by reducing intermediate feature map sizes, removing bypass connections, and pruning computational blocks, and \textit{ii}) \textit{hyperparameter tuning} optimizing the number of recurrent inferences within the model. 
We validate our tiny models on three real-world SoA datasets: MVSEC~\cite{zhu_multi_2018}, HKU~\cite{chen_esvio_2023}, and RPG~\cite{rpg}. 
Among many TinyDEVO configurations, our best-performing one achieves an 11.5$\times$ reduction in memory footprint and a 29.7$\times$ reduction in operations per frame compared to DEVO, requiring only \SI{63.8}{\mega\byte} and \SI{5.2}{\giga MAC/frame}.
With these reductions, TinyDEVO achieves a competitive ATE of 
\SI{27}{\centi\meter}, \SI{45.3}{\centi\meter}, and \SI{4.9}{\centi\meter} 
on MVSEC, HKU, and RPG, respectively, while processing real-world trajectories of up to \SI{100}{\meter}.
Compared to the SoA DEVO baseline, scoring an ATE of \SI{8.3}{\centi\meter}, \SI{25.9}{\centi\meter}, and \SI{0.9}{\centi\meter} on MVSEC, HKU, and RPG, respectively, our results are at most only \SI{20}{\centi\meter} higher across all datasets.

Finally, we deploy TinyDEVO on GAP9~\cite{ceredaTrainingFlyOnDevice2024}, a RISC-V parallel ULP System-on-Chip (SoC), where it achieves an energy consumption of \SI{79}{\milli\joule} per inference and an average power consumption of \SI{86}{\milli\watt} at \SI{370}{\mega\hertz}, including off-chip RAM memory.
The end-to-end execution reaches \SI{1.2}{frame/\second}, demonstrating for the first time the feasibility of a cutting-edge SoA event-based VO pipeline, running entirely on a sub-\SI{100}{\milli\watt} ULP embedded vision SoC.

\section{Related Work} \label{sec:related_work}

This section provides an overview of monocular VO algorithms, emphasizing event-based methods and energy-efficient pipelines. 
A summary of representative monocular approaches is reported in \Cref{tab:related_works}.

\noindent\textbf{Geometric vs. DL-based VO.}  
Traditional VO methods, such as SVO~\cite{forster_svo_2014}, DSO~\cite{engel2016directsparseodometry}, and ORB-SLAM3~\cite{orb_slam3}, rely on frame-based RGB cameras and geometric pipelines to reconstruct motion through feature extraction, matching, and 3D optimization.  
These approaches are computationally demanding, typically requiring high-end CPUs or GPUs with power budgets of 30–\SI{250}{\watt}~\cite{legittimoBenchmarkAnalysisDatadriven2023}, which makes them unsuitable for low-power embedded platforms.
Moreover, they generally achieve lower accuracy and robustness than modern DL-based methods~\cite{teed_deep_2023,legittimoBenchmarkAnalysisDatadriven2023}, which leverage learned feature representations.
For these reasons, we focus on DL-based VO and provide a comparison with geometric VO pipelines in Section~\ref{sec:discussion}.  

\noindent\textbf{Event-based VO.}  
Event-based VO pipelines have recently surpassed RGB-based methods~\cite{klenk_deep_2023,pellerito_rampvo_2024}, showing to be robust in challenging visual conditions~\cite{pellerito_rampvo_2024}.
Several approaches enhance motion estimation by leveraging additional sensing modalities, such as event-based stereo vision~\cite{Zhou_2021} or fusion with depth sensors~\cite{depth_event_odometry}.
However, such sensor fusion increases power consumption,  system complexity, and calibration overhead, making it unsuitable for ULP embedded hardware.
In this work, we therefore focus on DL-based monocular event-only VO, which is better suited for resource-constrained platforms and can optionally be complemented with an inertial measurement unit to improve robustness~\cite{deio, guan_pl-evio_2023}.

Among monocular event-based approaches, Zhu~\textit{et al.}~\cite{zhu_unsupervised_2018} and Ye~\textit{et al.}~\cite{ye2019unsupervisedlearningdenseoptical} trained convolutional neural networks (CNNs) to jointly predict camera pose, optical flow, and depth from event representations using the MVSEC dataset~\cite{zhu_multi_2018}.  
Despite improvements over RGB-based methods, these two works exhibit poor generalization, as they fail on indoor sequences and lack evaluation beyond the MVSEC dataset. 
The current state of the art in event-only VO is DEVO~\cite{klenk_deep_2023}, which adapts the RGB-based DPVO architecture~\cite{teed_deep_2023} to event-based inputs.  
DEVO demonstrates strong generalization, outperforming other event-based VO algorithms~\cite{rebecq_evo_2017, guan_pl-evio_2023} across seven real-world datasets.  
However, DEVO requires over \SI{733}{\mega\byte} of memory and \SI{155}{\giga \op/frame}, relying on powerful GPUs ($\sim$\SI{250}{\watt)} for achieving real-time inference.  
Thus, SoA monocular event-based VO has been demonstrated so far only on powerful processors.
In contrast, our work aims to design a lightweight, event-only VO algorithm suitable for deployment on resource-constrained ULP MCUs.

\noindent\textbf{Energy-efficient VO.}  
Energy-efficient monocular VO systems have so far been limited to RGB pipelines, often relying on application-specific integrated circuits (ASICs).
K\"uhne~\textit{et al.}~\cite{kuhne_low_2025} presented an embedded visual-inertial odometry (VIO) system, exploiting an ASIC for optical-flow computation and an ARM Cortex-A72 for VIO processing.
However, they achieve an average power consumption above \SI{3.7}{\watt}.
Suleiman~\textit{et al.}~\cite{suleiman_navion_2019} and Mandal~\textit{et al.}~\cite{mandal_visual_2019} proposed more energy-efficient solutions by designing dedicated ASIC accelerators for VIO, achieving average power consumption as low as \SI{2}{\milli\watt} while operating at \SI{30}{\fps} and \SI{20}{\fps}, respectively.
However, these ASIC-based designs are tailored to specific algorithms and lack flexibility for general-purpose workloads.
On general-purpose MCUs, Palossi~\textit{et al.}~\cite{palossi_vo} demonstrated an RGB-based VO algorithm running at \SI{117}{\fps} and \SI{10}{\milli\watt}, though it tackles basic hovering functionality and lacks validation with real-world data.
To the best of our knowledge, event-based monocular VO has not yet been demonstrated on ULP MCUs.
Our work addresses this gap by introducing a DL-based, monocular event-only VO algorithm designed for general-purpose ULP MCUs, enabling visual perception within a sub-\SI{100}{\milli\watt} power envelope.
Furthermore, we validate the proposed VO system across three real-world datasets to assess generalization.
\section{System Design and Optimization} \label{sec:system_design}

\begin{figure*}[!t]
    \centering
    \includegraphics[width=\linewidth]{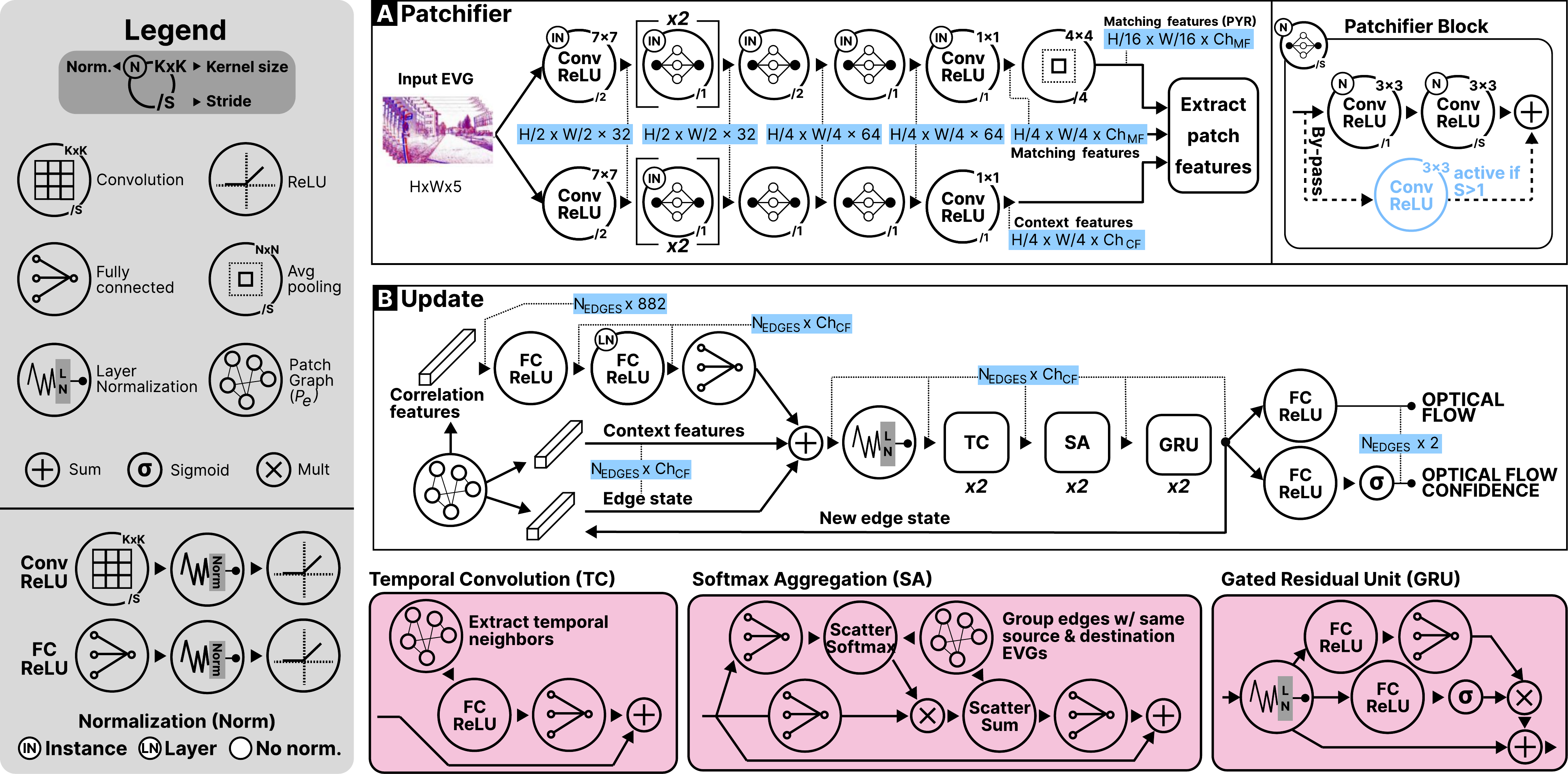}
    \caption{Diagram of DEVO's computational blocks optimized in this work: A) the \textit{patchifier}, and B) the \textit{update}.}
    \label{fig:devo_block_scheme}
    \vspace{-2mm}
\end{figure*}

\subsection{Background: DEVO} \label{subsec:devo_overview}

DEVO~\cite{klenk_deep_2023} takes as input a sequence of five event voxel grids (EVGs)~\cite{klenk_deep_2023, zhu_unsupervised_2018}, where raw events are accumulated into timestamped 2D event-frames.
To estimate camera poses, EVGs are processed through four stages: the \textit{patchifier}, the \textit{correlation} block, the \textit{update} block, and the \textit{bundle adjustment}.
The patchifier, detailed in \Cref{fig:devo_block_scheme}-A, is a CNN with two branches composed of convolutional layers and by-pass connections. 
It outputs two tensors: the \textit{matching features} ($MF$) and the \textit{context features} ($CF$).
The former consists of two tensors of sizes $\frac{W}{4} \times \frac{H}{4} \times Ch_{MF}$ and $\frac{W}{16} \times \frac{H}{16} \times Ch_{MF}$.
The latter (CF) is composed of one tensor of size $\frac{W}{4} \times \frac{H}{4} \times Ch_{CF}$.
In the original implementation of DEVO (i.e., the baseline), $Ch_{MF}$ and $Ch_{CF}$  are 128 and 384, respectively.

The correlation block processes MF from the current and past EVGs to produce compact $1\times882$ \textit{correlation features} that encode camera motion over a temporal window, i.e., the \textit{removal window} ($R_w$).
This block samples $N_{patches}=96$ tensors of size $3\times3\times Ch_{MF}$, referred to as \textit{patches}, from each MF within the last $R_w$ timestamps.
Each sampled patch becomes a node in the \textit{patch graph} ($\mathcal{P}_e$), where edges correspond to pairwise dot-products between patches and $7\times 7 \times Ch_{MF}$ tensors of an MF whose timestamps lie within a fixed temporal span called \textit{patch lifetime} ($P_{LT}$).
Consequently, each such dot-product yields an output correlation feature, and all MF within $R_w + P_{LT}$ timestamps must be retained in memory.

The update block, illustrated in \Cref{fig:devo_block_scheme}-B, is the most computationally demanding stage of DEVO.
It is a recurrent graph neural network that processes the correlation features and CFs, performing one forward pass for each edge in $\mathcal{P}{e}$.
Under the baseline configuration $(N_{patches}, R_w, P_{LT})=(96,22,13)$,  the total number of edges is 47712, computed as:
{ 
\small 
\begin{equation} 
\label{eq:n_edges} 
\begin{aligned} 
N_{\text{edges}} &= N_{patches}\!\left((R_w\!+\!1)P_{LT} + \frac{m(2(R_w\!+\!1)-1-m)}{2}\right),\\ m &= \min\{P_{LT}-1,\; R_w\}. \end{aligned} 
\end{equation} }
The update block consists of:
\textit{i}) two temporal convolutions (TCs), using fully connected (FC) layers to combine features from edges with adjacent timestamps, 
\textit{ii}) two softmax aggregations (SAs), which use scatter-softmax operations to combine features across edges connected either to the same patch or MF,
\textit{iii}) two gated residual units (GRUs) that process the input tensors with FC layers, ReLUs, a sigmoid, and a by-pass connection, 
\textit{iv}) two FC layers predicting optical flow and a confidence score.

Lastly, the update block's outputs are fed to a differentiable bundle adjustment that jointly optimizes camera poses and patch depths over a temporal \textit{optimization window} ($W_{opt}$).
This closes the loop between local correlations and global trajectory consistency.
The resulting poses from the bundle adjustment are used to prune edges in $\mathcal{P}e$ corresponding to negligible camera motion. 

We trained DEVO for \SI{180}{\kilo\nothing} iterations on four NVIDIA GH200 GPUs using the full TartanAir dataset~\cite{wangTartanAirDatasetPush2020}
As in~\cite{klenk_deep_2023}, we trained DEVO and all our netowrks using their custom loss functions, with a batch size of 1, $N_{patches}=80$, and employed the ATE as the evaluation metric.

\subsection{Ultra-low-power Hardware Platform}

The target ULP MCU we use in this work is the GWT GAP9 SoC~\cite{ceredaTrainingFlyOnDevice2024}.
GAP9 features two frequency domains: the Fabric Controller (FCtrl) with a single RISC-V core, and the Cluster (CL) with nine general-purpose RISC-V cores, four mixed-precision floating-point units (FPUs) (FP16/BF16/FP32), and the NE16 accelerator for \texttt{int8} $3\times3$ and $1\times1$ convolutions.
All CL cores support single instruction, multiple data (SIMD) execution: a 4-lane 8-bit integer SIMD on the cores and a 2-lane 16-bit SIMD on the FPUs.  
GAP9 integrates \SI{128}{\kilo\byte} of shared L1 scratchpad and \SI{1.5}{\mega\byte} of L2 SRAM; L2 accesses from the CL incur in $\sim$100 extra cycles of latency.
The GAP9 evaluation board provides $>$\SI{8}{\mega\byte} of external L3 HyperRAM.

Two DMAs manage L3-L2 and L2-L1 transfers, achieving, respectively \SI{370}{\mega\byte/\second} and \SI{13.3}{\giga\byte/\second} throughput.
DMAs enable efficient overlapping between memory transfers and computation, effectively masking L2 access latency in the case of compute-bounded workloads.
All experiments use GAP9 at its maximum frequency, i.e., \SI{370}{\mega\hertz}@\SI{0.8}{\volt}.
We use GWT’s GAP\textit{flow} framework to quantize and generate C code for the DL-based parts of our pipeline.
We pair GAP9 with the Prophesee GENX320 event-camera, featuring a resolution of $320\times320$\SI{}{\pixel} and a power consumption of 3--\SI{9}{\milli\watt}.

\subsection{DEVO Architecture Optimization}

To address the large memory and computational requirements of DEVO, we introduce several optimizations aimed at reducing both of them while maintaining ATE scores close to those of the original model.
Our optimizations focus on: architectural modifications on both the patchifier (\Cref{fig:devo_block_scheme}-A) and the update block (\Cref{fig:devo_block_scheme}-B), and reducing the number of edges in $\mathcal{P}e$. 

\noindent\textbf{Patchifier block.}
We optimize the size of the largest tensors in the algorithm, i.e., MF and CF, by reducing their number of channels ($Ch_{MF}$ and $Ch_{CF}$), thereby decreasing both the memory requirements and the computation needed in the correlation and update blocks.
Lowering $Ch_{CF}$ reduces the number of MAC operations more than reducing $Ch_{MF}$, since it decreases the input dimensionality of the update block, which is executed once per edge, while the reduction on $Ch_{MF}$ only impacts the final convolutional layer of the patchifier.

The baseline patchifier also produces two MF outputs that must be stored in memory for the last $R_w + P_{LT}$ EVGs and processed during each inference. 
This design inflates both peak memory and total operations in the patchifier and update blocks.  
To address this, we analyze the effect of removing the smaller of the two MF tensors, called PYR, with dimensions $\frac{W}{16} \times \frac{H}{16} \times Ch_{MF}$, consequently halving the size of the \textit{correlation features}.
Finally, we assess the removal of by-pass connections from the patchifier.
This modification is effective for compressing small CNNs~\cite{lamberti_pulpdronetv3, lamberti_tiny-pulp-dronets_2022}, simplifying their deployment on MCU-constrained devices, while slightly reducing the number of MAC operations with negligible drops in their accuracy.

\noindent\textbf{Update block.}
We evaluate three architectural modifications on the update block.
First, we investigate the removal of the TC and SA blocks.
While Teed \textit{et al.}~\cite{teed_deep_2023} report that combining TC and SA yields marginal ATE improvements for RGB-based VO, their effectiveness in event-based pipelines has not been verified.
Removing the TC blocks primarily reduces the number of operations by eliminating two FC layers, whereas removing the SA blocks drastically decreases both memory usage and computational cost.
The SA’s softmax operation represents a significant bottleneck for embedded deployment. 
In the baseline DEVO, \SI{37}{\mega\nothing} elements are processed per forward pass, and the computation for each softmax element ($e_{\sigma}$) requires about 380 cycles on a 16-bit FPU~\cite{belanoFlexibleTemplateEdge2025}.
Finally, we replace the GRU units, introduced initially to avoid vanishing gradients~\cite{teed_deep_2023}, with a lightweight alternative consisting of a normalization layer followed by two FC layers with a ReLU activation in between, decreasing the number of operations required.

\noindent\textbf{Patch graph optimization.}
DEVO’s inference latency is dominated by the total number of edges (\Cref{eq:n_edges}), as the correlation, update, and bundle adjustment blocks process each one of them.
Consequently, we run an ablation study over $(N_{patches}, W_S, P_{LT})$ hyperparameters to identify the best trade-off between memory footprint, MAC operations, and the resulting ATE.
\section{Experimental Results} \label{sec:results}

We evaluate our VO pipeline on three widely used event-based datasets: MVSEC~\cite{zhu_multi_2018}, HKU~\cite{chen_esvio_2023}, and RPG~\cite{rpg}. 
These datasets cover complementary operating conditions and sensing setups, providing a representative benchmark for real-world event-based VO~\cite{klenk_deep_2023,deio, depth_event_odometry,zhu_multi_2018,pellerito_rampvo_2024}. 
In particular, they span different trajectory scales, with average lengths of \SI{31.23}{\meter} for MVSEC, \SI{68.12}{\meter} for HKU, and \SI{10.5}{\meter} for RPG. 
They are also recorded using different event camera models and resolutions. 
MVSEC and HKU use a DAVIS346 sensor with resolution 346$\times$\SI{260}{\pixel}, while RPG is captured using a DAVIS240 sensor with resolution 240$\times$\SI{180}{\pixel}. 

Following~\cite{klenk_deep_2023}, we evaluate only the indoor sequences of MVSEC. 
To ensure stable evaluation, we trim the trajectories of MVSEC and HKU to remove EVGs generated while the camera is stationary (e.g., before take-off and after landing), where events only represent noise and inflate the ATE variance.
Specifically, we remove the first and last \SI{20}{\centi\meter} of each MVSEC trajectory and \SI{10}{\centi\meter} for HKU, which correspond to less than 5\% and 1\% of each sequence, respectively. 
Without this preprocessing step, the baseline DEVO~\cite{klenk_deep_2023} exhibits a $1.6\times$ increase in ATE on MVSEC, while the ATE of our VO models on MVSEC and HKU degrades by up to $2\times$ and $1.2\times$, respectively. 
As monocular VO produces trajectories with an unknown scale, we applied Umeyama alignment to the ground truth before evaluation as in~\cite{klenk_deep_2023}.

\subsection{DEVO Architecture Exploration}
\label{subsec:devo_architecture_exploration}

In this section, we evaluate the effect of incremental architectural modifications, as described in \Cref{sec:system_design}, on the two main building blocks of DEVO (\Cref{fig:devo_block_scheme}): \textit{i}) the patchifier, and \textit{ii}) the update.
We evaluate the models in terms of \textit{i}) peak memory footprint (Peak M.),   
\textit{ii}) operations (MACs and number of $e_{\sigma}$),
\textit{iii}) average ATE with its standard deviation ($\sigma$). 
The ATE is computed as in~\cite{klenk_deep_2023}: for each dataset, we evaluate the VO algorithm five times per sequence, take the median ATE across the five runs, and then average these medians over all sequences in the dataset.
Because the three evaluation datasets differ in trajectory length by up to an order of magnitude, we also report the average ATE normalized by sequence length ($\overline{\text{nATE}}$), defined as:
\begin{equation}
\label{eq:nATE}
\overline{\text{nATE}} = 
\frac{1}{|\mathcal{D}|} 
\sum_{d \in \mathcal{D}} 
\frac{1}{|\mathcal{S}_d|} 
\sum_{s \in \mathcal{S}_d} 
\frac{\text{ATE}_s}{L_s}
\end{equation}
where $\mathcal{D}=\{\text{MVSEC}, \text{HKU}, \text{RPG}\}$, $\mathcal{S}_d$ is the set of sequences in a dataset $d$, and $L_s$ the trajectory length of a sequence $s$.

\noindent\textbf{Channel shrinking.} 
This evaluation is reported in \Cref{tab:results.shrinking1}, where we explore different configurations of $Ch_{MF}$ and $Ch_{CF}$.
The baseline DEVO~\cite{klenk_deep_2023} ($Ch_{MF}=128$, $Ch_{CF}=384$) scores an ATE of \SI{8.3}{\centi\meter}, \SI{25}{\centi\meter}, and \SI{0.9}{\centi\meter} on MVSEC, HKU, and RPG, respectively.
It requires \SI{154.7}{\giga MACs}, $e_{\sigma}$=\SI{36.6}{\mega\nothing}, and a \SI{733}{\mega\byte} memory footprint. 
Halving the $Ch_{MF}$ to 64 does not affect the total number of MAC and $e_{\sigma}$, but it reduces the peak memory by \SI{7.3}{\percent} (\SI{53.5}{\mega\byte}), with only a minor ATE increase (\SI{1.1}{\centi\meter} at most on HKU).
$Ch_{MF}=64$ and $Ch_{CF}=192$ further lowers ATE by 0.3-\SI{3}{\centi\meter}, depending on the dataset, while drastically reducing memory by \SI{39}{\percent} (\SI{283}{\mega\byte}), MACs by \SI{67}{\percent} (\SI{104}{\giga MACs}), and $e_{\sigma}$ by 2$\times$ (\SI{18.3}{\mega\nothing}).
Further halving $Ch_{CF}$ to 96 yields even lower requirements---\SI{338}{\mega\byte} memory, \SI{19.7}{\giga MACs}, and $e_{\sigma}=\SI{9.1}{\mega\nothing}$---while increasing the ATE only marginally (+\SI{1.7}{\centi\meter} on MVSEC and +\SI{0.6}{\centi\meter} on RPG).
The smallest model ($Ch_{MF}=64$, $Ch_{CF}=96$) scores an ATE of \SI{13.6}{\centi\meter}, \SI{29.6}{\centi\meter}, and \SI{2.1}{\centi\meter} on MVSEC, HKU, and RPG, respectively.
Overall, shrinking $Ch_{MF}$ and $Ch_{CF}$ substantially reduces computational and memory requirements, achieving 7.9$\times$ fewer MACs, 4$\times$ fewer $e_{\sigma}$, and a 2.2$\times$ smaller memory footprint compared to the baseline~\cite{klenk_deep_2023}, with only a minor ATE increase of 1.2–\SI{5.3}{\centi\meter}.

\noindent\textbf{Patchifier by-pass removal.} 
Building upon the smallest configuration in \Cref{tab:results.shrinking1} (i.e., $Ch_{MF}=64$, $Ch_{CF}=96$), we study, in \Cref{tab:results.shrinking2}, the effect of removing the patchifier by-pass connections.
Without by-pass, the ATE improves slightly across all datasets (the error decreases of 0.2--\SI{0.8}{\centi\meter}), while memory and operations remain unchanged.  
We therefore adopt a model without by-pass connections for the following experiments.

\begin{table}[tb]
\centering
\caption{ATE comparison varying the number of channels of the matching features ($Ch_{MF}$) and correlation features ($Ch_{CF}$). 
}
\label{tab:results.shrinking1}
\setlength{\tabcolsep}{3pt} 
\resizebox{\linewidth}{!}{
\begin{tabular}{ccccc ccc} 
\toprule
\multicolumn{2}{c}{\textbf{Nº channels}} & 
\multicolumn{3}{c}{\textbf{Model}} & 
\multicolumn{3}{c}{\textbf{Avg. ATE [cm] / $\sigma$ ($\downarrow$)}} \\
\cmidrule(lr){3-5}
\cmidrule(lr){6-8}
$Ch_{MF}$ & $Ch_{CF}$ & \makecell{Peak M.\\{[}MB]} & \makecell{Params\\{[}M]} & \makecell{MACs\\{[}G]} & MVSEC & HKU & RPG \\ 
\midrule
128 & 384 & 733 & 3.39 & 154.7 & 8.3 / 2.2 & 25.9 / 40.1 & 0.9 / 0.3 \\
64  & 384 & 679 & 3.39 & 151.1 & 9.0 / 2.8 & 27.0 / 38.2 & 1.3 / 0.5 \\
128 & 192 & 504 & 1.22 & 51.2 & 12.1 / 1.4 & 32.4 / 37.1 & 1.6 / 0.4 \\
64  & 192 &  450  & 1.21 & 47.7  & 11.9 / 3.7 & 30.0 / 39.1 & 1.5 / 0.4 \\
\textbf{64} & \textbf{96} &  \textbf{338}   & \textbf{0.62} & \textbf{19.7} & 
\textbf{13.6 / 2.3} & \textbf{29.6 / 37.3} & \textbf{2.1 / 0.6} \\
\bottomrule
\end{tabular}%
}
\end{table}
\begin{table}[tb]
\caption{Impact of the Patchifier by-pass removal on the ATE.}
\label{tab:results.shrinking2}
\centering
\setlength{\tabcolsep}{3pt} 
\resizebox{\linewidth}{!}{
\begin{tabular}{@{}c c c c c c c c@{}} 
\toprule
\multicolumn{2}{c}{\textbf{Nº channels}} & 
\textbf{Patchifier} & 
\multicolumn{2}{c}{\textbf{Model}} & 
\multicolumn{3}{c}{\textbf{Avg. ATE [cm] / $\sigma$ ($\downarrow$)}} \\
\cmidrule(lr){4-5}
\cmidrule(lr){6-8}
$Ch_{MF}$ & $Ch_{CF}$ & by-pass &  \makecell{Peak M.\\{[}MB]} & \makecell{MACs / $e_{\sigma}$\\{[}G] / {[}M]} & MVSEC & HKU & RPG \\
\midrule
64 & 96 & yes & 338 & 19.7 / 9.2 & 13.6 / 2.3 & 29.6 / 37.3 & 2.1 / 0.6 \\
\textbf{64} & \textbf{96} & \textbf{no} & \textbf{338} & \textbf{19.7 / 9.2} & \textbf{13.0 / 2.8} & \textbf{28.4 / 35.8} & \textbf{1.9 / 0.6} \\
\bottomrule
\end{tabular}
}
\end{table}

\begin{table}[t]
\caption{Ablation study on the update block's architectural components: PYR, TC, SA, GRU. 
} 
\label{tab:results.shrinking3}
\centering
\footnotesize
\setlength{\tabcolsep}{2pt}
\begin{tabular}{@{}c c c c c c c c c c@{}}
\toprule
\multicolumn{4}{c}{\textbf{Update block}} & 
\multicolumn{2}{c}{\textbf{Model}} & 
\multicolumn{3}{c}{\textbf{Avg. ATE [cm] ($\downarrow$)}} &
\textbf{$\overline{\text{nATE}}$} ($\downarrow$) \\
\cmidrule(lr){5-6}
\cmidrule(lr){7-9}
TC & PYR & SA & GRU & 
\makecell{Peak M.\\ {[}MB]} & 
\makecell{MACs / $e_{\sigma}$\\{[}G] / {[}M]} &
MVSEC & HKU & RPG & \\ 
\midrule
\cmark & \cmark & \cmark & \cmark & 337.7 & 19.7 / 9.2 & 13.0 & 28.4 & 1.9 & 0.42 \\
\cmark & \cmark & \cmark &        & 337.7 & 18.8 / 9.2 & 13.0 & 30.0 & 4.3 & 0.52 \\
\cmark & \cmark &        & \cmark & 337.5 & 17.0 / 0.0 & 15.5 & 39.4 & 2.8 & 0.52 \\
\cmark & \cmark &        &        & 319.1 & 16.2 / 0.0 & 16.4 & 36.4 & 4.2 & 0.56 \\
\cmark &        & \cmark & \cmark & 250.2 & 15.9 / 9.2 & 14.8 & 32.7 & 4.4 & 0.55 \\
\textbf{\cmark} &        & \textbf{\cmark} &        & \textbf{250.2} & \textbf{15.0 / 9.2} & \textbf{14.7} & \textbf{33.4} & \textbf{2.2} & \textbf{0.47} \\
\cmark &        &        & \cmark & 250.0 & 13.3 / 0.0 & 15.6 & 46.4 & 4.2 & 0.59 \\
\cmark &        &        &        & 231.7 & 12.4 / 0.0 & 23.0 & 45.1 & 6.4 & 0.79 \\
       & \cmark & \cmark & \cmark & 337.6 & 17.9 / 9.2 & 53.4 & 41.4 & 4.5 & 1.30 \\
       & \cmark & \cmark &        & 337.5 & 17.0 / 9.2 & 101.3 & 56.6 & 5.5 & 1.95 \\
       & \cmark &        & \cmark & 337.4 & 15.3 / 0.0 & 109.4 & 52.2 & 12.4 & 2.59\\
       & \cmark &        &        & 300.6 & 14.4 / 0.0 & 96.4 & 53.3 & 17.6 & 2.56 \\
       &        & \cmark & \cmark & 250.1 & 14.2 / 9.2 & 41.5 & 45.7 & 3.4 & 1.20 \\
       &        & \cmark &        & 250.0 & 13.3 / 9.2 & 97.7 & 59.4 & 6.3 & 2.08 \\
       &        &        & \cmark & 249.9 & 11.5 / 0.0 & 152.1 & 59.1 & 26.5 & 3.65 \\
       &        &        &        & 213.2 & 10.6 / 0.0 & 143.4 & 58.4 & 24.3 & 3.39 \\
\bottomrule
\end{tabular}
\end{table}

\begin{figure*}[t]
    \centering
    \includegraphics[width=1\linewidth]{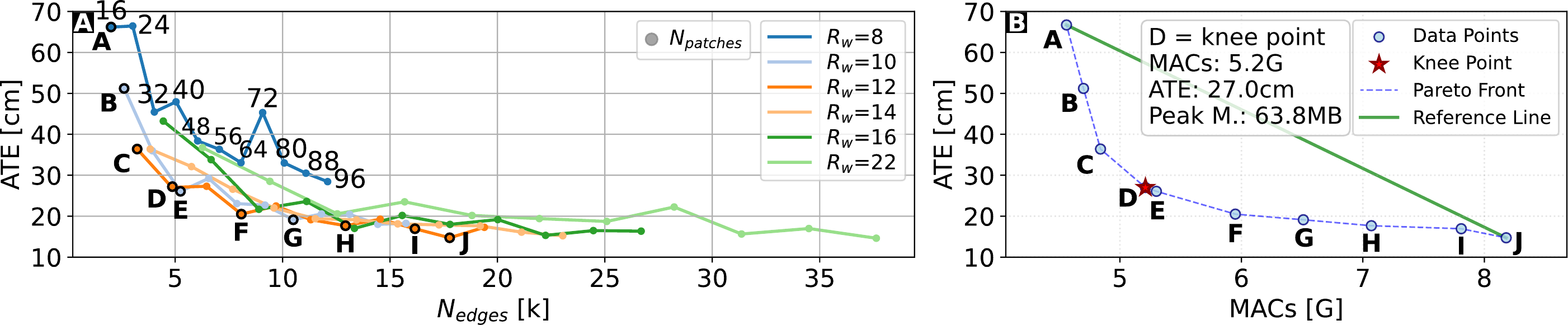}
    \caption{A) Graph optimization, sweeping $R_w$ (lines) and $N_{patches}$ (scatter points);  B) knee point on the Pareto front (A-J), finding the best trade-off model.}
    \label{fig:results_hyperparameters}
    \vspace{-2mm}
\end{figure*}

\noindent\textbf{Update block architecture.} 
Incrementally to previous architectural changes, in \Cref{tab:results.shrinking3} we evaluate the effects of removing the TC, PYR, SA, and GRU blocks.  
Removing TC does not reduce peak memory allocation; however, it severely degrades performance by penalizing the temporal correlation between patches, which is key for accurate trajectory reconstruction in event-based VO as evidenced by a 2.2$\times$ to 6.5$\times$ (4.3$\times$ on average) increase in $\overline{\text{nATE}}$ across all datasets.
Then, we consider the effect of PYR, SA, and GRU, while keeping TC active, by comparing row pairs that differ only in the presence of a single block.  
Removing PYR marginally increases $\overline{\text{nATE}}$ by 1.2$\times$ on average but saves \SI{87.5}{\mega\byte} and \SI{3.8}{\giga MACs} per frame.
Removing SA has a more relevant effect, as it aggregates information over many edges, producing coarser features that contribute to the VO robustness.
Its removal increases $\overline{\text{nATE}}$ by 1.3$\times$ on average, while saving \SI{2.7}{\giga MACs} and \SI{9.2}{\mega\nothing} $e_{\sigma}$. 
Removing GRU has the smallest impact, increasing $\overline{\text{nATE}}$ by only 1.1$\times$ on average while saving \SI{0.9}{\giga MACs}.

Overall, evaluating the presence of PYR, SA, or GRU produces comparable $\overline{\text{nATE}}$ degradation, yielding to different trade-offs in terms of memory footprint, MACs, and $e_{\sigma}$.
Among all these combinations, the best ATE is achieved by keeping TC and SA and removing PYR and GRU
Compared to the baseline update block (first row in \Cref{tab:results.shrinking3}), this increases $\overline{\text{nATE}}$ by only +0.06 while reducing memory by 26\%, to \SI{250}{\mega\byte}), and MACs by 24\%, to \SI{15.0}{\giga MACs}. 
We select for the next evaluation the model combining all previous architecture optimizations: $Ch_{MF}=64$, $Ch_{CF}=96$, no by-pass connections, no GRU, and no PYR.

\subsection{Reduction of Edges in the Patch Graph}

Building on the optimized architecture presented in \Cref{subsec:devo_architecture_exploration},
we analyze how reducing $N_{\text{edges}}$ affects the ATE.
As defined in \Cref{eq:n_edges}, $N_{\text{edges}}$ depends on three parameters, which we sweep as follows: $P_{LT} \in [8, 13]_{\text{step}=1}$, $N_{\text{patches}} \in [16,96]_{\text{step}=8}$, and $R_w \in$ \{8, 10, 12, 14, 16, 22\}.
The number of edges is also pruned at runtime (see \Cref{subsec:devo_overview}), but for this analysis, we assume the worst-case scenario in which no edges are removed.
We evaluate on MVSEC, as its low standard deviation in testing ATE ($\sigma<\SI{3.7}{\centi\meter}$) ensures reliable comparisons, and its average trajectory length (\SI{31.23}{\meter}) lies between HKU (\SI{68.12}{\meter}) and RPG (\SI{10.5}{\meter}), offering a balanced benchmark.

First, we set $P_{LT}$ to 10, as we observe that the ATE remains stable compared to the baseline value $P_{LT}=13$, while setting $P_{LT}$ to lower values leads to at least a 1.4$\times$ degradation in performance.
Then, in \Cref{fig:results_hyperparameters}-A, we compare ATE as a function of $N_{\text{edges}}$.
Line colors indicate different $R_w$ values, while the markers indicate increasing values of $N_{\text{patches}}$ from left to right along each curve, shown on $R_w=8$.
A consistent trend emerges: ATE increases as $N_{\text{patches}}$ (and thus $N_{\text{edges}}$) decreases.
Compared to the TinyDEVO baseline parameters ($R_w=22$, $N_{\text{patches}}=96$, $N_{\text{edges}}=38$k, ATE=\SI{14.7}{\centi\meter}), reducing $N_{\text{edges}}$ to about \SI{10}{\kilo\nothing} increases ATE to \SI{20}{\centi\meter} on average; the main exception is $R_w=8$, where ATE is \SI{30}{\centi\meter} or worse.
Below \SI{10}{\kilo\nothing} edges, all configurations show a steeper (convex) degradation up to \SI{65}{\centi\meter}.
In \Cref{fig:results_hyperparameters}, the Pareto-optimal points along the ATE and $N_{\text{edges}}$ trade-off curve are annotated with labels A-J.

As $N_{\text{edges}}$ is directly proportional to MACs (and thus latency), we analyze the Pareto points in \Cref{fig:results_hyperparameters}-B as ATE vs. MACs.
Model J achieves the best accuracy (\SI{14.7}{\centi\meter}) at \SI{8.2}{\giga MACs}, with $e_{\sigma}=\SI{3.4}{\mega\nothing}$ and \SI{108}{\mega\byte} peak memory, whereas the lowest-latency model A requires only \SI{4.6}{\giga MACs} with $e_{\sigma}=\SI{0.4}{\mega\nothing}$ and \SI{50.7}{\mega\byte}, but reaches \SI{66.7}{\centi\meter} ATE.
To select the best trade-off between ATE and MACs, we compute the \textit{knee point} geometrically: we draw a reference line from the first to the last Pareto points, and pick the point with the largest perpendicular distance to this line.
According to this criterion, the best trade-off is point D ($R_w{=}12$, $N_{\text{patches}}{=}24$), which reduces MACs by \SI{65}{\percent}, $e_{\sigma}$ by \SI{90}{\percent}, and peak memory by \SI{75}{\percent} relative to the baseline hyperparameters, while increasing ATE by only \SI{12.3}{\centi\meter} on MVSEC.

We call TinyDEVO the final model that combines the architectural optimizations ($Ch_{MF}{=}64$, $Ch_{CF}{=}96$, no by-pass, no GRU, no PYR) with the selected hyperparameters ($R_w=12$, $N_{\text{patches}}=24$, $P_{LT}=10$).
TinyDEVO achieves an ATE of \SI{27.0}{\centi\meter}, \SI{45.3}{\centi\meter}, and \SI{4.9}{\centi\meter} on MVSEC, HKU, and RPG, respectively, with \SI{5.2}{\giga MACs}, $e_{\sigma}=\SI{0.93}{\mega\nothing}$, and \SI{63.8}{\mega\byte} peak memory.
Compared to baseline DEVO~\cite{klenk_deep_2023}, TinyDEVO yields an average ATE that is about $3.5\times$ higher, but reduces MACs by $29.7\times$, $e_{\sigma}$ by $39.4\times$, and peak memory by $11.5\times$.
\Cref{fig:4.2-trajectories} shows example trajectories produced by TinyDEVO on the three datasets.
In the Supplementary material section we show real-time predictions of TinyDEVO, and the ground truth trajectory, on one sequence of the MVSEC dataset.


\subsection{Inference and Power Profiling on GAP9}

\begin{figure*}[t]
    \centering
    \includegraphics[width=1\linewidth]{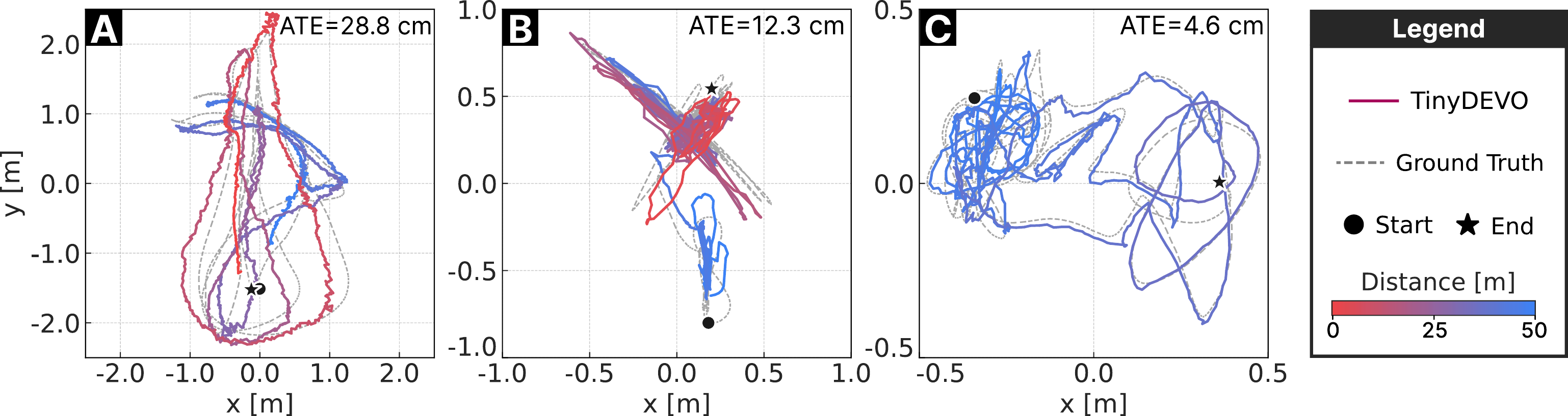}
    \caption{Sample trajectories predicted by TinyDEVO across three datasets: A) MVSEC, B) HKU, and C) RPG.}
    \label{fig:4.2-trajectories}
    \vspace{-2mm}
\end{figure*}

\begin{table}[tb]
\centering
\caption{Latency of DEVO vs. TinyDEVO on the GAP9 MCU.}
\label{tab:results.inference}
\setlength{\tabcolsep}{2pt}
\scriptsize
\resizebox{\columnwidth}{!}{
\begin{tabular}{cccccccccc}
\toprule
\multirow{2}{*}{\textbf{Model}} & 
\multirow{2}{*}{\textbf{Input [\SI{}{\pixel}]}} &
\multirow{2}{*}{\textbf{$N_{\text{edges}}$}} &
\multicolumn{5}{c}{\textbf{Latency [s]}} &
\multirow{2}{*}{\textbf{FPS}} \\
\cmidrule(l){4-8}
& & & \textbf{PATCH} & \textbf{CORR} & \textbf{UPD} & \textbf{BA} & \textbf{TOT} & \\
\midrule
\multirow{2}{*}{\makecell{DEVO~\cite{klenk_deep_2023}\\ (\texttt{fp16}/\texttt{int8})}}
& $346\times260$ & 47712 & 0.18 & 4.92 & 39.76 & 0.14 & 45,00 & 0.02 \\
& $240\times180$ & 47712 & 0.08 & 4.92 & 39.76 & 0.14 & 44.90 & 0.02 \\
\midrule
\multirow{2}{*}{\makecell{\textbf{TinyDEVO}\\ (\texttt{fp16}/\texttt{int8})}}
& $346\times260$ & 4848 & 0.15 & 0.39 & 0.35 & 0.04 & 0.93 & 1.1 \\
& $240\times180$ & 4848 & 0.06 & 0.39 & 0.35 & 0.04 & 0.85 & 1.2 \\
\bottomrule
\end{tabular}
}
\end{table}

\begin{figure}[b]
    \includegraphics[width=\linewidth]{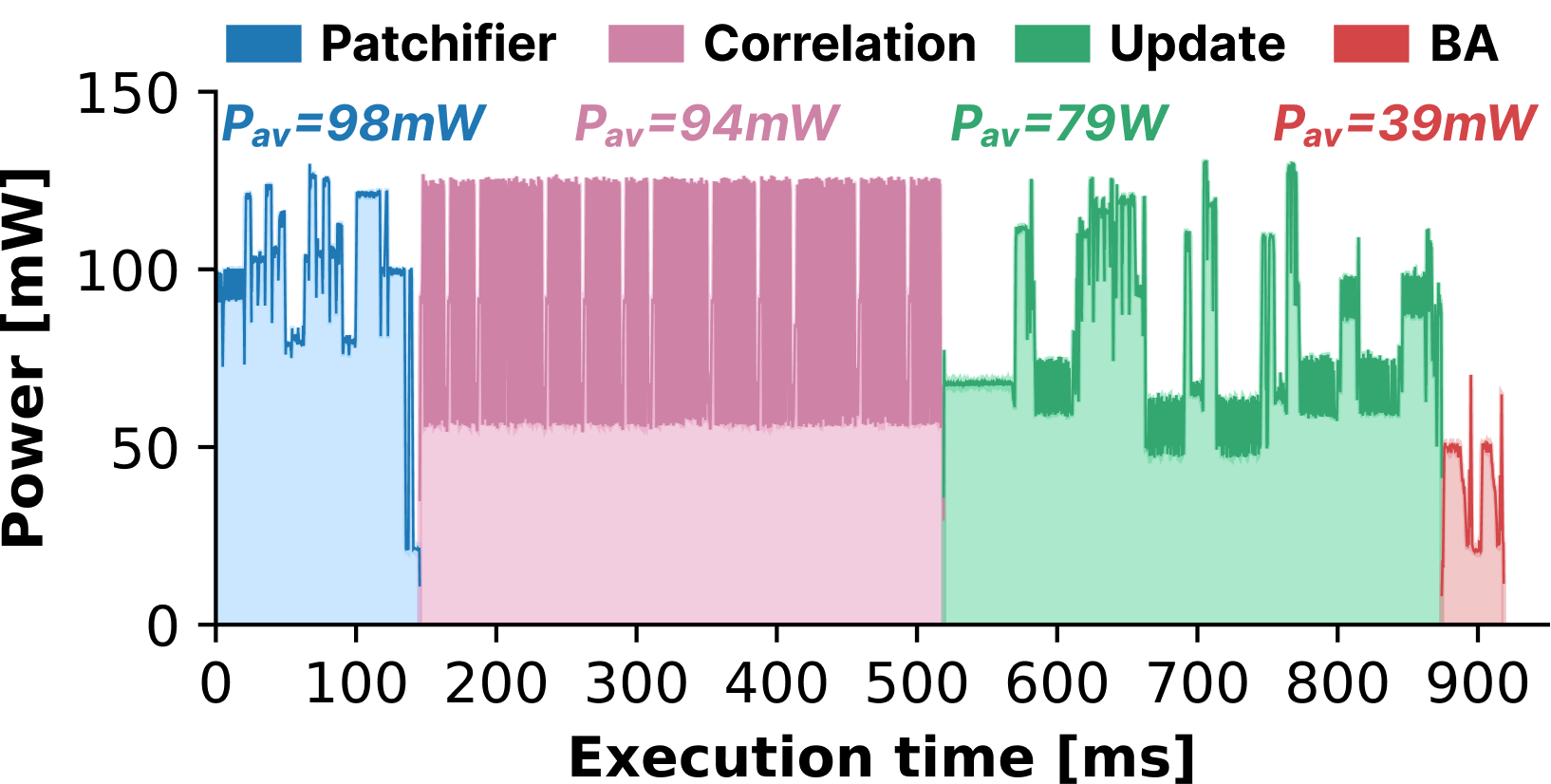}
    \caption{Power waveforms of the GAP9 EVK at FCtrl@\SI{370}{\mega\hertz}, CL@\SI{370}{\mega\hertz}, Vdd@\SI{0.8}{\volt} executing TinyDEVO, using 346$\times$\SI{260}{\pixel} inputs.}
    \label{fig:power}
\end{figure}

We deploy our TinyDEVO on the GAP9 SoC, and profile its execution latency and power consumption.
We quantize the DL-based blocks, i.e.,  patchifier (PATCH) and update (UPD), to \texttt{int8}, and the geometric blocks, i.e., correlation (CORR) and bundle adjustment (BA), to \texttt{FP16} and \texttt{BF16}, respectively, following prior works that show negligible increases in numerical error~\cite{jacobQuantizationTrainingNeural2018,float16,float16_2}.
With this mixed-precision quantization scheme, the peak memory footprint using an input size of $346\times260$\SI{}{\pixel} is \SI{252}{\mega\byte} for DEVO and \SI{26.1}{\mega\byte} for TinyDEVO.

All measurements are performed with the GAP9 at \SI{370}{\mega\hertz}, $V_{\!dd}=\SI{0.8}{\volt}$ (both FCtrl and CL). 
\Cref{tab:results.inference} reports the latency of DEVO and TinyDEVO at two input resolutions: $346\times260$\SI{}{\pixel} (MVSEC, HKU) and $240\times180$\SI{}{\pixel} (RPG).
On $346\times260$\SI{}{\pixel} inputs, TinyDEVO is faster than DEVO with a per-block speedup of $\sim$$1.22\times$ (PATCH), $\sim$$12.6\times$ (CORR), $\sim$$114\times$ (UPD), and $\sim$$3.2\times$ (BA).
Using $240\times180$\SI{}{\pixel} inputs affects only the execution time of the PATCH, and TinyDEVO, vs. DEVO, yields a speedup of $1.24\times$, which leads to an end-to-end speedup of $48\times$ with $346\times260$\SI{}{\pixel} inputs and $53\times$ when using an input size of  $240\times180$\SI{}{\pixel}.

Finally, we measure TinyDEVO’s power consumption using a Nordic Semiconductor Power Profiler II and the GAP9 evaluation board (EVK).
The power waveforms (\Cref{fig:power}) account for both SoC and off-chip L3 HyperRAM power consumption, excluding the event-camera.
The PATCH, CORR, UPD, and BA blocks have an average power consumption of \SI{98}{\milli\watt}, \SI{94}{\milli\watt}, \SI{79}{\milli\watt}, and \SI{39}{\milli\watt}, respectively.
The PATCH and UPD blocks exhibit the highest power consumption peaks as they are executed on NE16 and perform frequent L3 memory accesses, which alone draw approximately $\sim$\SI{60}{\milli\watt}.
The CORR block and the BA stage consume less power because they run on the 9 cores of the CL; the former requires L3 access to fetch MFs, while the latter can rely exclusively on the on-chip L2 memory.
Overall, our TinyDEVO runs on the GAP9 at 1.1-\SI{1.2}{\fps} within \SI{86}{\milli\watt}, corresponding to \SI{79}{\milli\joule} per frame.
To the best of our knowledge, we demonstrate, for the first time, a SoA event-based VO running on a ULP MCU within \SI{100}{\milli\watt},
making it compatible with the computing power budget of miniaturized robots~\cite{lamberti_pulpdronetv3,suleiman_navion_2019, bouwmeesterNanoFlowNetRealtimeDense2023, potocnikCircuitsSystemsEmbodied2024, ceredaTrainingFlyOnDevice2024} and smart glasses~\cite{bartoliLynXEventBasedGesture2025,freyGAPsesVersatileSmart2025}.

\subsection{Comparison vs. Geometric-based Approaches} \label{sec:discussion}

We compare the DL-based TinyDEVO with traditional geometric monocular VO approaches.
The SoA RGB-based geometric VO algorithm is ORB-SLAM3~\cite{orb_slam3}, which, relying on geometric feature extraction and optimization, achieves an ATE of \SI{2.97}{\centi\meter} on the RPG dataset while requiring approximately \SI{900}{\mega\byte} of peak memory~\cite{legittimoBenchmarkAnalysisDatadriven2023}. 
Considering event-based methods, the SoA geometric approach EVO~\cite{rebecq_evo_2017} achieves \SI{10.10}{\centi\meter} ATE on the same dataset~\cite{klenk_deep_2023}. 
Since EVO does not report memory consumption, we measured it using its open-source release: EVO reaches a peak memory usage of \SI{534.7}{\mega\byte} at a resolution of 240$\times$\SI{180}{\pixel} on the RPG dataset.

In contrast, our TinyDEVO achieves \SI{2.2}{\centi\meter} ATE while requiring only \SI{64}{\mega\byte} of peak memory. 
This corresponds to a $1.35\times$ improvement in accuracy and a $14\times$ reduction in memory compared to ORB-SLAM3~\cite{orb_slam3}, and a $4.5\times$ improvement in accuracy with an $8.4\times$ lower memory footprint compared to EVO~\cite{rebecq_evo_2017}.
These results demonstrate that our DL-based TinyDEVO achieves a significantly better accuracy-memory trade-off than both RGB- and event-based geometric VO pipelines, while remaining suitable for memory-constrained embedded platforms.

\section{Conclusion} \label{sec:conclusion}

We presented TinyDEVO, an event-only, DL-based monocular VO model tailored for ultra-low-power MCUs.
Compared to the SoA DEVO, TinyDEVO reduces memory by $11.5\times$ and operations by $29.7\times$, with only a \SI{19}{\centi\meter} increase in average trajectory error.
Through targeted architectural optimizations and hyperparameter tuning, we reduce the footprint to \SI{63.8}{\mega\byte} and \SI{5.2}{\giga MACs/frame}.
Running on a 9-core RISC-V ULP MCU, TinyDEVO achieves \SI{1.2}{frame/\second} at just \SI{86}{\milli\watt}.
On the one hand, this result marks a soft real-time performance that, to the best of our knowledge, represents the first demonstration of a SoA event-based VO algorithm running on ULP MCUs.
On the other hand, our contribution is essential in paving the way toward high-throughput hard real-time VO pipelines for ULP processors.

\section*{Acknowledgment}
This work was partially supported by the SNSF RoboMix2 project (grant nb. 10004854) and by the Swiss National Supercomputing Centre under project IDs lp12 and lp160.

{
    \small
    \bibliographystyle{ieeenat_fullname}
    \bibliography{bibliography}
}



\end{document}